# Nonlinear control of PT-symmetry and non-Hermitian topological states

Shiqi Xia[1+], Dimitrios Kaltsas[2+], Daohong Song[1,3+], Ioannis Komis[2], Jingjun Xu[1,3], Alexander Szameit[4], Hrvoje Buljan[1,5], Konstantinos G. Makris[2,6] and Zhigang Chen[1,3,7]

[1]The MOE Key Laboratory of Weak-Light Nonlinear Photonics, TEDA Applied Physics Institute and School of Physics, Nankai University, Tianjin 300457, China
[2]Department of Physics, University of Crete, Heraklion, 71003, Greece
[3]Collaborative Innovation Center of Extreme Optics, Shanxi University, Taiyuan, Shanxi 030006, China
[4]Institut für Physik, Universität Rostock, Albert-Einstein-Strasse 23, 18059 Rostock, Germany
[5]Department of Physics, Faculty of Science, University of Zagreb, Bijenička c. 32, 10000 Zagreb, Croatia
[6]Institute of Electronic Structure and Laser (IESL) – FORTH, Heraklion, 71110, Greece
[7]Department of Physics and Astronomy, San Francisco State University, California 94132, USA
[+]These authors made equal contributions.

hbuljan@phy.hr, makris@physics.uoc.gr, zgchen@nankai.edu.cn

Advances in topological photonics and non-Hermitian optics have drastically changed our perception on how interdisciplinary concepts may empower unprecedented applications. Bridging the two areas could uncover the reciprocity between topology and non-Hermiticity in complex systems. So far, such endeavors have focused mainly on linear-optics regime. Here, we establish a nonlinear non-Hermitian topological platform for control of parity-time (PT) symmetry and topological edge states. Experimentally, we demonstrate that optical nonlinearity effectively modulates the gain and loss of a topological interface waveguide in a non-Hermitian Su-Schrieffer-Heeger lattice, leading to switching between PT and non-PT-symmetric regimes accompanied by destruction and restoration of topological zero modes. Theoretically, we examine the fundamental issue of the interplay between two antagonistic effects: the sensitivity close to exceptional points and the robustness of non-Hermitian topological modes. Realizing single-channel control of global PT-symmetry via local nonlinearity may herald new possibilities for light manipulation and unconventional device applications.

About a dozen years ago, two important concepts were severally introduced to the realm of photonics, namely, the quantum Hall edge state[1, 2] and the parity-time (PT) symmetry[3, 4], leading to the birth of two ever-thriving areas - topological photonics[5] and non-Hermitian optics[6, 7]. On the one hand, topologically protected edges states and photonic topological insulators were realized in a variety of platforms, including gyro-optic materials, helical waveguide arrays, aperiodic coupled resonators, bianisotropic metamaterials and synthetic crystalline photonic structures[5, 8-11]. On the other hand, by manipulating the role played by gain and loss, active and passive PT-symmetry in optics has also provided a plethora of alternative design platforms for unconventional control of light, aiming towards unique photonics devices based on non-Hermitian physics[6, 7, 12-14].

Intertwining these two different areas of photonics occurred naturally, leading to a new direction of non-Hermitian topological photonics where the interplay between non-Hermiticity and topology takes place. Indeed, several experiments have demonstrated topological nature of edge states in non-Hermitian systems, either with or without global PT-symmetry[15-18], although the existence of such topological states was initially debated. In fact, it has now been realized that non-Hermitian properties can give rise to unusual topological phenomena including for example unusual non-Hermitian topological light steering and funneling[19-24]. Perhaps, one of the most striking developments closest to technological applications is the realization of topological insulator lasers[25-27], in which topological photonics and non-Hermitian optics naturally coalesce and conspire: lasing is based on topologically protected modes and a laser system is inherently non-Hermitian due to presence of gain and loss. Topological lasers are found to exhibit superior features such as reduced lasing threshold, enhanced stability, and single-mode operation.

Notwithstanding the synergetic outcome of the two areas, much of the venture in non-Hermitian topological photonics has so far taken place mainly in the linear-optics regime. Apart from topological lasers which inherently involve nonlinearity, nonlinear effects like optical solitons were explored separately in the two different domains, focusing on either their topological[28-31] or non-Hermitian[32, 33] aspects. Moreover, although nonlinear phenomena exist in a variety of topological systems[34], many open questions remain unanswered with respect to how nonlinearity would change the dynamics in non-Hermitian topological systems. In particular, how can we characterize a non-Hermitian topological system driven by nonlinearity? Can PT-symmetry and topological states be manipulated solely by nonlinear control in non-Hermitian systems?

In this work, we demonstrate a scheme for single-channel nonlinear control of PT-symmetry and nonlinearity-induced restoration/destruction of non-Hermitian topological states. Our experimental platform is based on specific photonic Su-Schrieffer-Heeger (SSH)[35, 36] lattices consisting of cw-laser-writing continuous ("gainy") and sectioned ("lossy") waveguides and an interface defect (see Fig.1), yet the concept developed here applies to a broad spectrum of non-Hermitian systems that have intensity-dependent gain or loss. Counterintuitively, even though the optical nonlinearity changes only the real part of the refractive index of a bulk material, we find that it can be employed to manipulate both the real and imaginary parts of a waveguide potential. This leads to an active control of otherwise "lossy", "gainy" or "neutral" non-Hermitian SSH lattices, switching them between PT- and non-PT-symmetric regimes. We also analyze theoretically the effect of nonlinearity on the robustness of topological defect modes as well as the eigenvalue sensitivity of the zero mode around the exceptional point. Our work represents a first attempt for single-channel tuning of a complex system with underlying dynamics driven by the interplay among topology, non-Hermiticity, and nonlinearity.

**Scheme for single-channel nonlinear tuning of PT-symmetry**

It is well known that an "active" linear non-Hermitian PT-symmetric system can be directly mapped onto a system with only loss simply by introducing a global decay factor (equivalent to offset the imaginary part of the gain-loss profile)[6]. In such "passive" PT-symmetric systems, non-Hermitian PT phenomena has been demonstrated without employing actual gain[12, 17]. We thus propose a scheme for single-channel nonlinear tuning of PT-symmetry and topological states in a passive PT-symmetric SSH lattice, which can be readily realized in our experiment, as illustrated in Fig. 1. The SSH lattice represents a prototypical one-dimensional (1D) topological system with chiral symmetry[5], as has been popularly employed for the study of topologically protected quantum states[37, 38], nonlinearity-driven topological effects[39-43], and topological lasing[44-46]. Different from previous work, in which losses were introduced to achieve passive-PT symmetric systems by using femto-second laser-written wiggled or scattered waveguides[17, 47], or by depositing lossy metal stripes on top of silicon waveguides[22], here we employ direct cw-laser-writing technique[48] to establish non-Hermitian SSH lattices in a bulk nonlinear crystal. As shown in the left panels of Fig. 1, the continuous waveguides (marked in red) represent the "gainy" ones, and sectioned waveguides can be "lossy" (marked in blue) or "neutral" (marked in green) depending on the gap size introduced between sections. Details about how the loss is introduced in sectioned waveguides and judiciously controlled by nonlinearity can be found in Supplementary Note 2. With proper control of the sectioned waveguides, a passive PT-symmetric SSH lattice can be realized first (middle panel). Then, under the action of self-focusing nonlinearity experienced by a probe beam at the interface, it can turn into a passive non-PT "gainy" system (top panel), as self-focusing reduces diffraction loss and leakage (or equivalently provides "gain") in the center waveguide. Likewise, under the action of self-defocusing nonlinearity, it can turn into a passive non-PT "lossy" system (bottom panel), since now the nonlinearity enhances the leakage and thus entails more loss in the waveguide. In this way, single-channel nonlinearity can actually affect the whole lattice, leading to switching between a PT- and a non-PT-symmetric system. Since the three SSH lattices (PT-symmetric with a "neutral" defect, non-PT-symmetric with a "gainy" defect, and non-PT-symmetric with a "lossy" defect) can all be created initially by laser-writing, such dimerized lattices provide a convenient platform to achieve nonlinearity-induced switching between PT- and non-PT-symmetric phases, thereby to explore the dynamics of topological states in the non-Hermitian system. Interestingly, the Hamiltonians of these three different non-Hermitian lattices are inherently related (see Fig 1). Such an underlying connection directly affects the corresponding complex eigenvalue spectra across the exceptional point as analyzed below.

**The non-Hermitian SSH model with nonlinearly controlled interface**

The lattices illustrated in Fig. 1 can be considered as two semi-infinite SSH dimer chains connected by a topological defect at the interface. For theoretical analysis, let us examine the topological states in a non-Hermitian *active* SSH system with a dimerization defect, as illustrated in Fig. 2(a). Under the tight-binding approximation, the dynamics of the system is governed by the following set of coupled mode equations[16, 17]:

$$-i\frac{\partial}{\partial z}\varphi_n = \beta^*\varphi_n + c_1\varphi_{n-1} + c_2\varphi_{n+1}, \quad n = 2, 4, \ldots \ldots \text{ or } -1, -3, \ldots \ldots \quad (1a)$$

$$-i\frac{\partial}{\partial z}\varphi_n = \beta\varphi_n + c_2\varphi_{n-1} + c_1\varphi_{n+1}, \quad n = 1, 3, \ldots \ldots \text{ or } -2, -4, \ldots \ldots \quad (1b)$$

$$-i\frac{\partial}{\partial z}\varphi_0 = \beta_0\varphi_0 + c_2\varphi_1 + c_2\varphi_{-1}, \quad n = 0 \quad (1c)$$

where $\varphi_n$ denotes the modal optical field amplitude in the $n$-th waveguide, $\beta = \alpha + i\gamma$ ($\alpha$ is the real part of the waveguide potential, and $\gamma$ is the imaginary part representing gain or loss), $c_1$ and $c_2$ are the strong and weak coupling coefficients, respectively, and $\beta_0$ denotes the potential of the defect waveguide at $n = 0$. If $\gamma = 0$ for all waveguides, the SSH lattice returns to the well-known Hermitian model that can support topologically protected mid-gap (zero-mode) states[39]. Even when the loss/gain is introduced ($\gamma \neq 0$), the non-Hermitian SSH lattice described above can still support a PT-symmetric topological interface state provided that there is no gain or loss at the dimerization defect[17], i.e., $\beta_0 = \alpha$, $\gamma_0 = 0$. Assuming that the lattice is terminated at the weak-coupling bond ($c_2$) so no edge states present on either ends[36], we summarize the results in Fig. 2(b) to show how an interface state is affected by non-Hermiticity and nonlinearity. The above coupled mode equations can be expressed in a convenient matrix form, and the relations between the non-Hermitian matrix Hamiltonians $\mathcal{H}_G, \mathcal{H}_L$, and $\mathcal{H}_N$ (corresponding to lattices with a "gainy", "lossy" and "neutral" interface defect) are given in Fig. 1 (see Supplementary Note 4 for details).

In the linear regime, $\alpha$ is the same for all waveguides, and a typical PT-symmetric topological interface state located right at the middle of the gap is illustrated by point $A$ in Fig. 2(b), where the left panels show the two-band diagram and eigenvalues of the lattice, and the right panels are the corresponding mode profiles. For all the calculations of Fig. 2, the coupling coefficients are taken as $c_1 = 4$, $c_2 = 1$, and the lattice consists of 16 waveguides in each side of the interface. (Note that from now on the linear propagation constant for all waveguides is set as $\alpha = 0, \gamma = 1$ except for the center one at $n = 0$). As seen from the top-left panel of Fig. 2(b), all eigenmodes have only real eigenvalues, since the lattice respects the PT-symmetry (unbroken regime)[17]. In the nonlinear regime, the propagation constant of a given waveguide potential is generally intensity-dependent, i.e., $\beta(I) = \alpha(I) +$

$i\gamma(I)$. As such, the eigenvalue of the topological state can be moved away from the mid-gap by nonlinearity, as shown already for the Hermitian SSH system[36, 43]. In accordance with our experimental situation, a probe beam initially excites only the center defect channel while it experiences an overall loss in the passive non-Hermitian lattice. As such, it is reasonable to consider that the nonlinearity is present only in the single channel in the center, therefore we have: $\beta_0(I) = \alpha_0(I) + i\gamma_0(I)$, where $I$ is the intensity of the excitation beam. When the nonlinearity only changes the real part of the potential while keeping $\gamma_0 = 0$, the eigenvalue of the zero-mode is shifted away from the center of the gap, moving upward (or downward) due to the self-focusing (or -defocusing) effect. These scenarios correspond to the modes marked by $B$ (or $C$) in Fig. 2(b), where $\beta_0$ is set at 2 (or $-2$). Clearly, the eigenmode profiles (shown in the right panels) remain symmetric as that of the mid-gap mode $A$, because the lattice overall still preserves the PT-symmetry. By contrast, if the nonlinearity changes the imaginary part of the potential $\gamma_0$, the PT-symmetry of the SSH lattice is destroyed. To simulate these scenarios, $\beta_0$ is set to $2i$ (or $-2i$), and the corresponding results are marked by $D$ (or $E$) in Fig. 2(b). In this case, the imaginary part of eigenvalues is shifted away from the zero-mode position, indicating that the non-Hermitian lattice is no longer PT-symmetric. Noticeably, in this non-PT regime, the eigenmode profiles become asymmetric with respect to the center defect, as more energy of the mode goes to the "lossy" ($D$) or "gainy" ($E$) waveguides depending on the sign of the nonlinearity. Therefore, by nonlinear excitation of the defect channel in the SSH lattice, observation of asymmetrical mode profiles also serves as a signature for the change of the imaginary part of the waveguide potential, indicating whether the PT-symmetry is present or not. This provides the guidance for our experiments.

**Experimental realization of the non-Hermitian SSH lattices**

To demonstrate the nonlinear tuning of PT-symmetry as illustrated in Fig. 1, we need to establish sectioned waveguides to obtain the desired non-Hermitian SSH lattices. The simple experimental setup is sketched in Fig. 3(a), where a stripe beam from a cw-laser is employed to write the waveguides in a biased SBN:61 photorefractive crystal. The waveguides are written sideways one by one[48], with either a uniform (continuous) stripe beam (for writing the

"gainy" waveguides) or a periodically modulated (sectioned) stripe beam (for writing the "lossy" and "neutral" waveguides). Shutters in Fig. 3(a) selectively control the entrance of the writing beam in path 1 and the probe beam in path 2. In the entire writing process, the bias field is $E_0 = 160 kV/m$ and the ordinarily-polarized writing beam has a power of about $200 \mu W$. Other experimental details are included in the Supplementary Note 1.

A passive PT-symmetric SSH lattice requires precise control of loss, so that $\gamma = 2\gamma_1$ in all "lossy" waveguides, $\gamma = \gamma_1$ in the center "neutral" waveguide, and $\gamma = 0$ in all "gainy" waveguides. To achieve the desired overall loss, the total number of waveguide sections (with section length $l$) in each channel is the same (here we have 17 sections in the 20-mm-long crystal), but spacing between adjacent sections (characterized by a gap length $m$) is smaller in the "neutral" waveguide as compared to that in all "lossy" waveguides. A superimposed writing beam pattern is shown in Fig. 3(b), consisting of alternating continuous and sectioned stripes except for the center defect channel. A typical SSH waveguide lattice written this way is shown in Fig. 3(c), which is examined by a broad beam (quasi-plane wave) as a probe sending along path 2. The strong and weak coupling corresponds to the smaller $(b = 15.4 \mu m)$ and larger $(a = 22.8 \mu m)$ channel separation, respectively, as marked in Fig. 3(b). The power transmission in three different ("gainy", "neutral" and "lossy") waveguides is in fact different due to different losses (see the insets in Fig. 3(d)). In Fig. 3(d), we plot the normalized intensity transmission ratio (defined as $y = I_{out}/I_0$, where $I_{out}$ and $I_0$ is the output intensity of the same probe beam from a sectioned and a continuous waveguide, respectively) as a function of the "gap ratio" (defined as $x = m/l$, which controls the waveguide loss). Clearly, as the gap length $m$ increases, the loss in the waveguide increases, thus the transmission decreases. The plot in Fig. 3(d) is obtained by applying a numerical beam propagation method based on the paraxial wave equation with a waveguide potential (see Supplementary Note 2 for detail), and the loss coefficient $\gamma$ is determined from the intensity transmission $I_{out} = I_0 \exp(-2\gamma L)$, where $L = 20 mm$ corresponds to the crystal length. This plot serves as a guideline for determining the parameters for the writing beams used in experiment, as shown in the three insets in Fig. 3(d). For example, the "gainy" waveguide ($\gamma = 0$) corresponds to the red dot at $(x, y) = (0, 1)$, since it is continuous ($m=0$) and lossless (neglecting Fresnel reflection and assuming the crystal has no absorption). The

"neutral" waveguide marked by the green dot at $(x, y) = (0.40, 0.70)$ corresponds to a gap ratio of 0.40 and a transmission ratio of 0.70, which yields $\gamma_1 = 8.93 m^{-1}$. From this, we can in turn find the parameters for the "lossy" waveguides, marked by the blue dot at $(x, y) = (0.56, 0.49)$, with a gap ratio of 0.56 to obtain the desired loss coefficient of $2\gamma_1$. The three insets in Fig. 3(d) are the outputs of a probe beam obtained in experiment, indicating a good agreement between experiment and simulation. Therefore, the SSH lattice established with such judicially designed writing beams fulfils the requirement for the PT-symmetry.

**Nonlinearity-induced transition from PT- to non-PT-symmetry**

Once the passive PT-symmetric SSH lattice is established in experiment (Fig. 3c), a cylindrically focused extra-ordinarily-polarized probe beam is sent into the "neutral" waveguide channel in the center (see illustration in the left of Fig. 4). When the probe beam undergoes linear propagation (i.e., without the bias field), a symmetric topological interface state corresponding to point $A$ in Fig. 2b is observed, as shown in Fig. 4(a3, b3), indicating that the non-Hermitian lattice in this linear case respects the PT-symmetry[17]. Conveniently, in the photorefractive crystal a self-focusing or defocusing nonlinearity can be achieved by applying a positive or negative electric field[39]. We now employ such a nonlinearity to demonstrate the PT transition graphically illustrated in Fig. 1.

We first fix the power of the probe beam at $2.5\mu W$ and set the bias field to be $-60 kV/m$ to introduce the self-defocusing effect. With the buildup of defocusing nonlinearity, the probe beam induces an anti-guide so that its energy escapes from the center defect channel. This equivalently introduces more leakage (loss) to the center waveguide (or imaginary part of the potential), turning the SSH lattice from passive PT-symmetric to non-PT-symmetric phase. As such, the excited mode becomes highly asymmetric in intensity distribution (Fig. 4(a4, a5)). Results shown in Fig. 4(a4, b4) correspond to those of point $E$ in Fig. 2(b), as more light goes to the "lossy" waveguide next to the center defect (see blue dots in Fig. 4(a3)). In contrast, when a self-focusing nonlinearity is employed (with a positive bias field of $100 kV/m$), it induces self-guiding of the probe beam so that its diffraction loss is suppressed, equivalently providing gain to the center waveguide. Again, the beam turns into asymmetric distribution shown in Fig. 4(a2, b2), correspond to point $D$ in Fig. 2(b), as more light goes to the nearby "gainy" waveguide. If the self-focusing nonlinearity is too strong, the beam becomes highly confined into the defect channel (Fig. 4(a1)), corresponding to a self-trapped nonlinear mode residing in the semi-infinite gap but not attributed to topological origin[43]. As emphasized before, the change in the real part of the index potential alone does not result in asymmetric modes. Therefore, these results represent the nonlinearity-induced transition from a PT-symmetric to a non-PT "gainy" or "lossy" system pictured in Fig. 1.

**Nonlinear restoration of PT-symmetric topological states**

As illustrated in Fig. 1, the transition from a PT to a non-PT lattice should be reversible by nonlinearity, which can be used for restoring the PT-symmetric topological states. Such an implementation is shown in Fig. 5, where two non-Hermitian SSH lattices are constructed by laser-writing with either a "gainy" (left) or a "lossy" (right) interface waveguide in the center. Let us consider Fig. 5(a) for example, the SSH lattice is initially equivalent to a non-PT "gainy" system (corresponding to the top panel in Fig. 1), so a probe beam evolves linearly into an asymmetric distribution (Fig 5(a3)). Under the nonlinear self-defocusing condition, however, the beam turns into a more symmetric profile with the characteristic feature of a topological mid-gap state: minimum amplitude in the two nearest-neighbor waveguides but non-zero amplitudes symmetrically distributed in the two next-nearest-neighbor waveguides (see point *A* in Fig. 2). This restoration of the topological state occurs due to equivalent loss which is introduced by the self-defocusing nonlinearity into the otherwise "gainy" waveguide in the center, entailing the retrieval of lattice PT-symmetry. If the strength of self-defocusing nonlinearity is too high (so the loss in center waveguide is beyond $\gamma_1$), the interface state becomes asymmetric again with more intensity going to the "lossy" channels (Fig. 5(a5), as the lattice falls into a non-PT "lossy" system. In the other direction with a self-focusing nonlinearity, the PT-symmetry cannot be restored, as the nonlinearity increases the gain-loss imbalance, eventually leading to a self-trapped state outside of the mid-gap [Fig. 5(a1, a2)], similar to that of Fig. 4(a1).

The scenario corresponding to an inversed transition starting from a non-PT "lossy" system is shown in Fig. 5(b), where an initially asymmetric interface state (Fig. 5(b3)) in the linear regime turns into a symmetric topological interface state (Fig. 5(b2)) as the self-focusing nonlinearity brings the non-Hermitian SSH lattice back to the PT-symmetric phase. Direct comparison of results in Fig. 5 and those in Fig. 4 supports clearly the nonlinear control of PT-symmetry and non-Hermitian topological interface states illustrated in Fig. 1, as also corroborated by our numerical simulations (see Supplementary Note 3).

**Discussion:**

Topology and PT-symmetry typically describe the global properties of a system, whereas most of the optical nonlinearities are local. Therefore, their interplay in some sense is a manifestation of the interplay of locality and globality. Despite the fact in our system nonlinearity only changes the real part of the refraction index (in the bulk of the material), we introduced here a method for constructing "passive" non-Hermitian lattices, relying on that nonlinearity can effectively control the loss of a waveguide, that is, the imaginary part of the waveguide potential. Our method provides an ideal platform to explore nonlinear effects in non-Hermitian topological systems, which so far have gone AWOL in photonic or any other experiments.

It is natural to ask: is there a general theory to study the non-Hermitian PT-symmetric systems driven by nonlinearity? In our current experiment and theoretical analysis, it is assumed that the modes experience nonlinearity only in the central defect waveguide, and we have focused on single-mode excitation and its propagation. To develop a general theoretical framework is beyond the scope of this paper; however, this can be done by extending the concepts introduced recently for nonlinear Hermitian topological systems[43]. We highlight the key idea here: Consider a dynamically evolving wavepacket in a nonlinear system whose linear counterpart is topological, non-Hermitian and PT-symmetric. The linear structure is described by a complex $z$-independent refractive index $\delta n_L$, whereas the nonlinear index $\delta n_{NL}$ depends on the amplitude of the beam and is $z$-dependent. Then, the dynamics of such a system should be governed by the nonlinear wave equation with an effective potential described by $\delta n_L + \delta n_{NL}$. By calculating its nonlinear eigenmodes and eigenvalues that evolve along the propagation axis $z$, one can identify the properties of the nonlinear system that are *inherited* from the corresponding linear system, thereby unravel *emergent* topological and non-Hermitian phenomena mediated by nonlinearity[43]. For instance, as analyzed below, the eigenvalues of the nonlinear modes are more robust to disorder (thus more stable) if the parameters are closer to the corresponding Hermitian topologically modes. The stability is therefore *inherited* from topologically protected zero-modes, even when such topological protection in SSH lattices is, strictly speaking, lost due to non-Hermicity and nonlinearity. Such an approach is somewhat analogous to the well-known KAM theorem addressing nonintegrable systems.

Another intriguing aspect is the study of nonlinear effects on the exceptional point (EP) - a special kind of eigenvalue degeneracy unique to non-Hermitian Hamiltonians[4, 7]. In particular, for some value of the control parameter (such as the global gain/loss amplitude) two or more eigenvalues and their corresponding eigenstates coalesce at the EP. Even though we cannot examine the behavior of our non-Hermitian lattices close to the EP due to experimental limitations, we highlight our theoretical finding here with respect to this point. In Fig. 6(a), we plot the range of the imaginary part of the complex eigenvalues (characterized by the eigenvalues with the maximal/minimal magnitudes of the imaginary part) as a function of the global gain/loss ratio of the waveguides. For the three lattices defined earlier, we keep the potential of the central defect waveguide fixed, as set by the strength of the nonlinearity. By varying the gain/loss ratio for all other waveguides, the bifurcation curves for the three lattices are dramatically different: Before the EP of the PT-symmetric lattice which has only real spectrum, the other two lattices have complex conjugate spectra which have different ranges of imaginary maximal/minimal magnitudes. Surprisingly, for some critical value of gain/loss ratio beyond the EP, all three lattices exhibit the same imaginary eigenvalues determined by the bulk modes of the lattices. This is in fact a direct outcome of the inherent connection between the Hamiltonians (Supplementary Note 4). At the EP of the PT symmetric system, the gap between the bands closes to zero and the topologically protected mode becomes extended; this indicates that a topological phase transition could have taken place. (Due to experimental limitations, the non-Hermitian lattices were fabricated for a particular global gain/loss ratio of the waveguides as illustrated in Fig. 6a, away from the EP).

It is well known that a non-Hermitian system exhibits enhanced sensitivity to external perturbations close to the EPs[49]. On the other hand, a direct outcome of the topological nature of the SSH model is the zero-mode robustness to off-diagonal perturbations. Therefore, another fundamental question arises: which of the two opposite tendencies will prevail, the sensitivity or the robustness? We theoretically address this question in Fig. 6(b), where the eigenvalues of the defect mode are plotted on the complex plane for various values of defect potential $\beta_0$, whereas the global gain/loss amplitude is fixed for the rest of the lattice. In other words, we examine the robustness of the defect mode to off-diagonal perturbations (on the coupling coefficients) that respect the lattice chiral symmetry. Strictly speaking, only the PT-symmetric lattice supports exactly the zero mode and, as a result, the complete topological robustness. Once the eigenvalue of the defect mode driven away from the central (0, 0) position in the complex plane, the topological protection is gradually lost. Interestingly enough, this loss of protection is not "isotropic" (in a sense that the instability of the defect mode grows in a preferred direction in the complex spectra), and it is enhanced when the global gain/loss amplitude is tuned close to the EP. Such novel effects certainly merit further investigation.

Still, for nonlinear non-Hermitian topological systems, there are many other fundamental questions yet to answer. For instance, how to characterize the topological invariants for finite non-Hermitian SSH lattices driven by nonlinearity, as one cannot simply employ the formulae for calculating the Chern number or the Zak phase developed for infinite Hermitian systems? More intriguingly, how would nonlinearity affect the topological phase and classification of symmetry and topology in non-Hermitian systems[50]? Undoubtedly, the interaction and synergy between nonlinearity, topology and non-Hermiticity will lead to a new paradigm for control of complex systems and for development of advanced photonic devices.

**Methods**

Methods, including statements of data availability and any associated accession codes and references, are available in the online version of this paper.

**Acknowledgements**

This research is supported by the National Key R&D Program of China under Grant No. 2017YFA0303800, the National Natural Science Foundation (11922408, 91750204, 11674180), PCSIRT, and the 111 Project (No. B07013) in China. H.B. acknowledge support in part by the Croatian Science Foundation Grant No. IP-2016-06-5885 SynthMagIA, and the QuantiXLie Center of Excellence, a project co-financed by the Croatian Government and European Union through the European Regional Development Fund - the Competitiveness and Cohesion Operational Programme (Grant KK.01.1.1.01.0004).

**Conflict of interests**

The authors declare no conflicts of interest. The authors declare no competing financial interests.

**Contributions**

All authors contributed to this work.

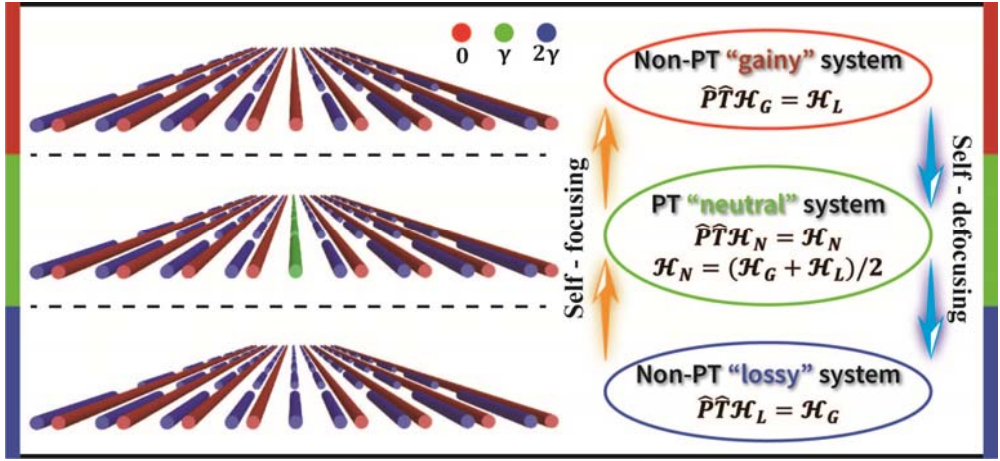

**Figure 1. Illustration of single-channel nonlinear tuning of PT-symmetry**. A "passive" PT-symmetric SSH lattice (middle panel) consisting of alternating continuous and sectioned waveguides can be switched to a non-PT "gainy" system (top panel) or a non-PT "lossy" system (bottom panel), thanks to the self-focusing or -defocusing nonlinearity along the topological defect channel at the center. The switching directions can be readily reversed, leading to destruction and restoration of the topological PT-symmetric interface state. In experiment, the cylinders illustrated here are replaced by one-dimensional continuous "gainy" (red) and sectioned "neutral" (green) or "lossy" (blue) waveguides via laser-writing, representing *passive* PT lattices. Red, green and blue dots represent "gainy", "neutral" and "lossy" lattice sites ($\gamma$ represents the imaginary part of a waveguide potential), and vertical colored bars denote that the corresponding non-Hermitian system is non-PT-symmetric "gainy", PT-symmetric "neutral" and non-PT-symmetric "lossy", respectively. The underlying relations between the matrix Hamiltonians that connect the three active non-Hermitian lattice models is also shown here, as analyzed in the Supplementary Note 4.

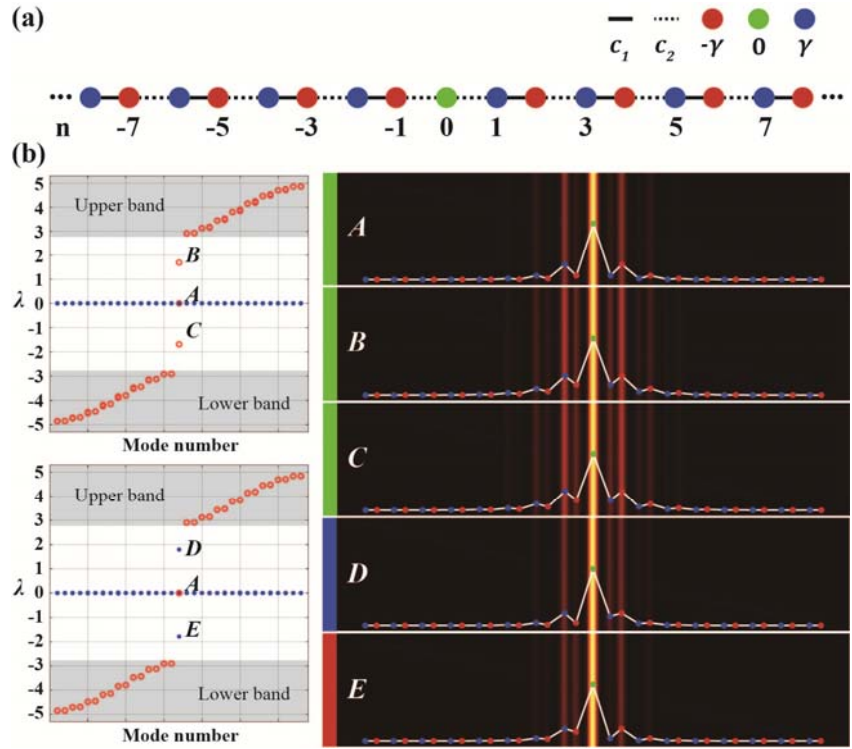

**Figure 2. The non-Hermitian SSH lattice and its topological interface states.** (a) Illustration of a PT-symmetric SSH lattice with an interface topological defect located at site $n$=0, where $c_1$ and $c_2$ denote the strong and weak coupling coefficients, and the colored dots represent different lattice sites. (b) Left panels show calculated eigenvalues for a finite lattice with 33 sites, where red circles and blue dots denote real and imaginary parts of the eigenvalues, and shaded regions illustrate the two-band structure of an infinite lattice. Right panels show the corresponding eigenmode profiles, where the eigenvalues for points $A$ to $E$ are obtained when the propagation constant $\beta_0$ of the center waveguide is changed to 0, 2, $-2$, $2i$ and $-2i$, respectively, while keeping $\beta^*$ for all the "gainy" and $\beta$ for all the "lossy" waveguides unchanged. Color codes for different waveguides and PT phases are the same as in Fig. 1.

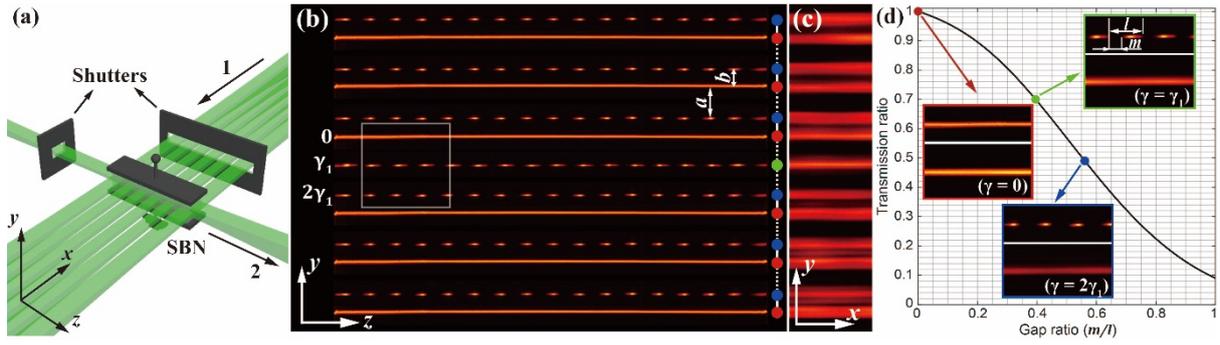

**Figure 3. Realization of a passive PT-symmetric SSH lattice with an interface**. (a) Schematic for cw-laser-writing of waveguides, where paths 1 and 2 are for writing and probing beams, respectively, launched into a 20 mm-long nonlinear SBN crystal. The shutters in each path control the entrance of the beams as needed. (b) Sideview of the writing beam pattern, where the gap between sectioned waveguides is different for the center "neutral" ($\gamma = \gamma_1$) and all other "lossy" ($\gamma = 2\gamma_1$) waveguides. The "gainy" waveguides ($\gamma = 0$) are all continuous. The waveguide spacing is $a = 22.8 \mu m$ and $b = 15.2 \mu m$, which determines the coupling for the dimer. (c) The written passive PT-symmetric lattice as examined by a broad plane-wave beam. (d) Plot of transmission ratio ($I_{out}/I_0$) as a function of the gap ratio ($m/l$) in a single waveguide from simulation. Three insets show the sideview of writing beams (top) and corresponding output patterns (bottom) of same probe beam from the gainy, neutral, and lossy waveguides taken in experiment, where desired losses are consistent with numerical simulation.

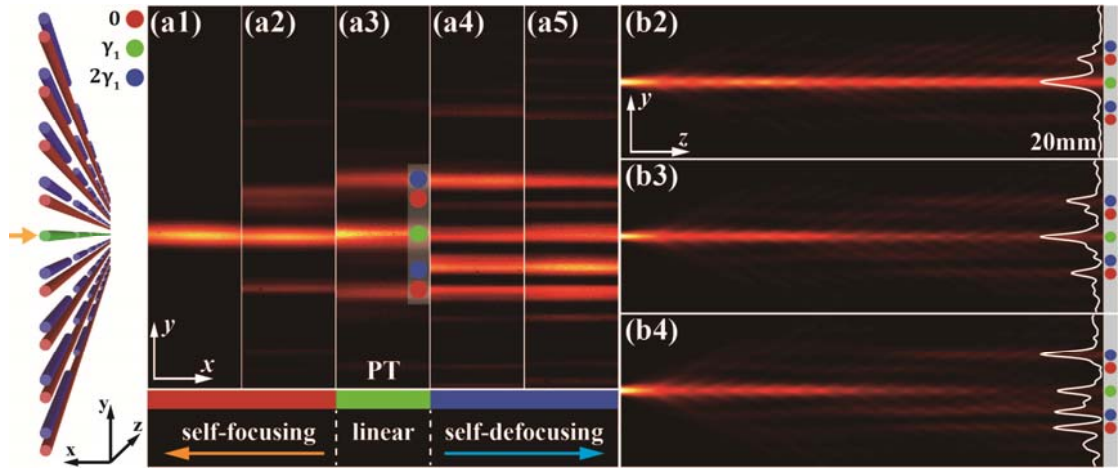

**Figure 4. Observed single-channel nonlinear destruction of a PT-symmetric topological interface state.** (a1-a5) Experimental results showing output transverse patterns of a probe beam launched into the center defect channel. (b2-b4) Simulation results showing sideview of propagation corresponding to (a2-a4). Left panel illustrates a non-Hermitian SSH lattice initially under "passive" PT-symmetry, which supports a linear PT-symmetric topological state (a3, b3). With increased strength of self-focusing nonlinearity, the lattice turns into a non-PT "gainy" system (as illustrated in Fig. 2), so the mid-gap state becomes asymmetric as more energy goes to the nearby "gainy" waveguides (a2, b2), corresponding to $D$ in Fig. 1(b). The situation for transition to a non-PT "lossy" system under self-defocusing nonlinearity is shown in (a4, b4), where more energy goes to the nearby "lossy" waveguides. This can be seen more clearly from superimposed intensity profiles at $z = 20mm$ (white lines) in (b2-b4). When the nonlinearity is too high, the beam becomes strongly localized or delocalized (a1, a5), corresponding to excitation of nonlinear modes not of topological origin.

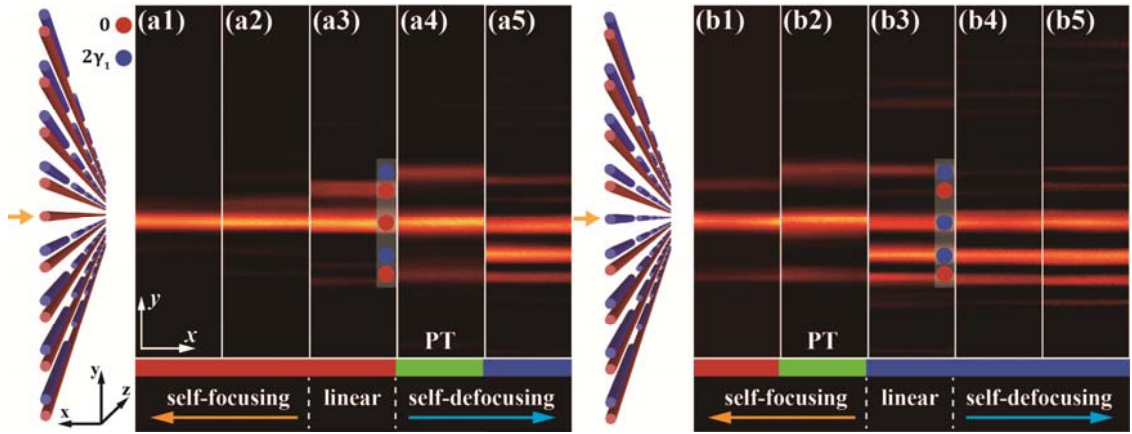

**Figure 5. Nonlinear restoring of topological interface states in an initially non-PT lattice.** The non-Hermitian SSH lattice is fabricated with a gainy (left) or lossy (right) interface waveguide, so it is initially at non-PT-symmetric phase. A probe beam launched into the center channel cannot evolve into a symmetric topological interface state in the linear regime (a3, b3), but a symmetric topological state is established under the action of self-defocusing (or -focusing) nonlinearity in the non-PT "gainy" (or "lossy") lattice, corresponding to results shown in (a4) (or (b2)). At different strength of nonlinearity, more localized or delocalized outputs of the probe beam are shown in (a1-a5) and (b1-b5), as corroborated by numerical simulation detailed in Supplementary Note 3.

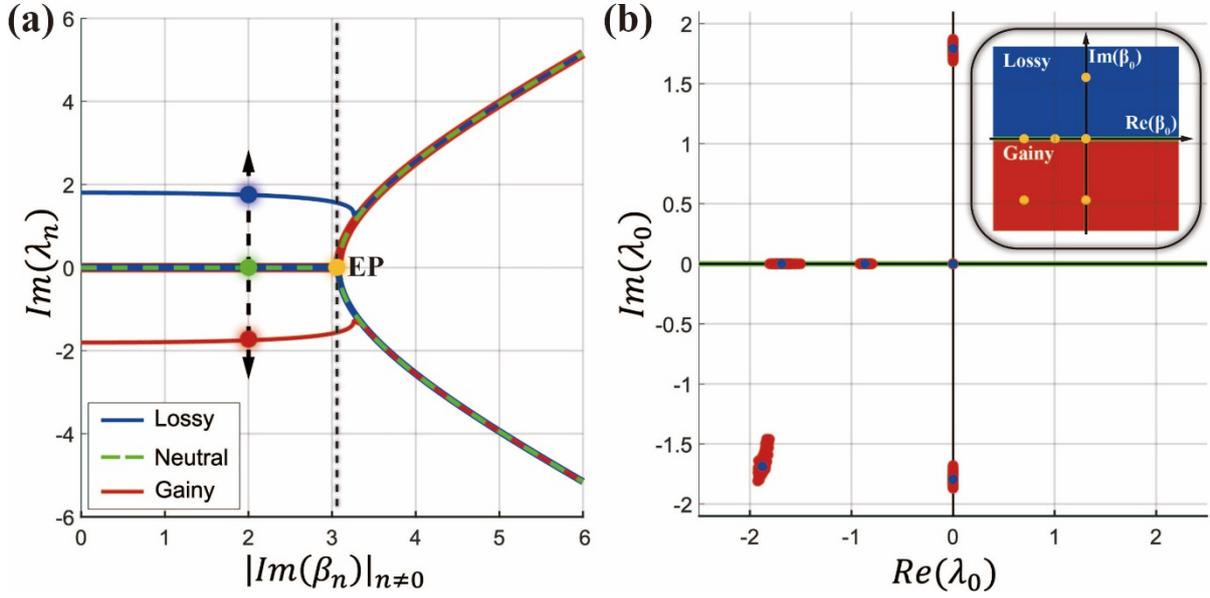

**Figure 6. Theoretical analysis of complex spectra around exceptional point and zero-mode robustness.** (a) The eigenvalue spectra of the three lattices (dashed-green for the "neutral", solid-blue for the "lossy", and solid-red for the "gainy") vs. global gain-loss amplitude ratio of the lattices. Plotted here are the eigenvalue envelopes formed by the maximal and minimal imaginary parts, while the values for the central defect potential are fixed in all three lattices. Exceptional point (EP) is marked for the "neutral" PT-symmetric lattice, beyond which the imaginary eigenvalues for the three lattices becomes identical. The three colored dots before the EP correspond to the experimental gain/loss parameters used for single-channel tuning ($\gamma_0 = 2i, 0, -2i$ for the "lossy", "neutral" and "gainy" systems, respectively). Other lattice parameters are the same as for Fig. 2 ($c_1 = 4$, $c_2 = 1$; the lattice has 33 sites). (b) Sensitivity of the defect mode eigenvalues to the perturbation on the coupling coefficients for a fixed gain-loss amplitude. The red dots denote the eigenvalue fluctuation from 100 realizations of added perturbation, and the blue dot corresponds to the initial defect eigenvalue. The inset graphically depicts the corresponding values of the central defect potential $\beta_0$ on the complex plane. Notice the perfect stability for the exact zero-mode at the origin in (b).